\documentclass[12pt]{article}%
\usepackage{heck}
\usepackage{cite}
\usepackage{graphicx}
\usepackage{makeidx}
\usepackage{multicol}
\usepackage{geometry}
\usepackage{amsfonts}
\usepackage{mathrsfs}
\usepackage{amssymb}
\usepackage{amsmath}%

\providecommand{\U}[1]{\protect\rule{.1in}{.1in}}
\numberwithin{equation}{section}

\newcommand{\bea}{\begin{eqnarray}}
\newcommand{\eea}{\end{eqnarray}}
\newcommand{\be}{\begin{equation}}
\newcommand{\ee}{\end{equation}}

\newcommand{\bem}{\begin{pmatrix}}
\newcommand{\eem}{\end{pmatrix}}

\def\U{\Upsilon}

\def\cn{{\cal N}}
\def\co{{\cal O}}

\begin{document}

\date{February, 2010}


\institution{SISSA}{\centerline{${}^{1}$Scuola Internazionale Superiore di Studi Avanzati via Beirut 2-4 I-34100 Trieste, ITALY}}

\institution{HarvardU}{\centerline{${}^{2}$Jefferson Physical Laboratory, Harvard University, Cambridge, MA 02138, USA}}%

\title{2d Wall-Crossing, R-twisting, and a Supersymmetric Index}
%

\authors{Sergio Cecotti\worksat{\SISSA}\footnote{e-mail: {\tt cecotti@sissa.it}}, and Cumrun Vafa\worksat{\HarvardU}\footnote{e-mail: {\tt vafa@physics.harvard.edu}}}%

\abstract{Starting from $\cn=2$ supersymmetric theories in 2 dimensions, we formulate a novel time-dependent supersymmetric quantum theory where the R-charge is twisted along the time.   The invariance of the supersymmetric
index under variations of the action for these theories leads to two predictions:
In the IR limit it predicts how the degeneracy of BPS states change as we cross the walls
of marginal stability.  On the other hand, its equivalence with the UV limit relates this
to the spectrum of the $U(1)$ R-charges of the Ramond ground states at the conformal point.
This leads to a conceptually simple derivation of results previously derived
using tt* geometry, now based on time-dependent supersymmetric quantum mechanics. }%

\maketitle

\tableofcontents

\pagebreak

\section{Introduction}
Supersymmetric quantum field theories often lead to exactly computable
quantities.  This arises from the fact that quantum corrections for many amplitudes
can be shown to vanish.  A well known example of this is the Witten index which
counts the net number of ground states of the theory.  With more supersymmetry
there are other quantities that can also be computed exactly.  For example with $\cn=2$ supersymmetry
in $d=2$ at the IR limit one can compute both the mass and the spectrum of BPS states, or
at the UV limit one can compute the spectrum of R-charges of Ramond ground states exactly.
In particular the masses of the BPS state interpolating between vacua $a$ and $b$  is given by 
$$m_{ab}=|Z_{ab}|$$
where $Z_{ab}\in \mathbb{C}$ is the central charge of the supersymmetric algebra in the $ab$ sector
(for an $\cn=2$ Landau-Ginzburg theory $Z_{ab}=W_a-W_b$ where $W_a$ denotes the value of the superpotential
at the critical point corresponding to the $a$-th vacuum).
The number of such BPS states is given by an integer $N_{ab}$ (which can be positive or negative, depending
on the fermion number of the kink). 

There is an interesting subtlety in this `exact' prediction of BPS masses and degeneracies:  As we change
the coupling constants of the theory it can happen that the phases of central charges of BPS
states can align.  Passing through such a configuration, the number of BPS states can jump.  In particular
suppose $Z_{ab}$ and $Z_{bc}$ align.  Then the number of BPS states in the $ac$ sector jumps according to
$$N_{ac}\rightarrow N_{ac}\pm N_{ab}N_{bc}$$
where the $\pm$ in the above formula depends on the orientation of the crossing of the phases.
This fact was derived in the case of LG theories by explicit computation \cite{classification} and can be explained
in full generality using the continuity of contribution of BPS
particles to certain computation as one crosses the wall \cite{fendleyetal}.

On the other hand one can consider the UV limit of such theories, where the BPS masses go to zero,
and we obtain a conformal theory.  One would naturally ask what is the conformal theory imprint
of the solitons and their degeneracies.  It was shown in \cite{classification}  by studying the solutions
to  tt* equations and comparing the corresponding monodromy data from UV to IR that the spectrum
of the $\cn=2$ $U(1)$  R-charges of the Ramond ground states are captured by the soliton degeneracies\footnote{An
alternative derivation of this result using D-branes was given in \cite{hiv}.}.
In particular for each sector $ab$
consider the operators  acting on the space of vacua given by
\begin{equation}\label{SINT}
	S_{ab}=I + N_{ab} T_{ab}
\end{equation}
where $I$ is the identity operator, and $T_{ab}$ is the `upper/lower triangular' operator which takes the $a$-th
vacuum to the $b$-th vacuum, and $N_{ab}$ is the net number of solitons in that sector. One considers
\begin{equation}
\label{phase}
M=T\Big(\prod_{ab} S_{ab}\Big)
\end{equation}
where the product is `T' ordered, in the sense of ordered according to phases of $Z_{ab}$.
Then
$${\rm eigenvalues\  of\  }M=\{ {\rm exp}(2\pi i q_a) \} $$
or, for $k$ integer,
\begin{equation}
I_k={\rm Tr}\  M^k=\sum_i {\rm exp}\big(2\pi i k q_a\big)
\label{maineq}
\end{equation}
where $q_a$ are the R-charges  (\textit{i.e.} $\cn=2$  $q_a={q_a}_L={q_a}_R$ charges) of the Ramond ground states at the conformal point.\footnote{For an $\cn=2$ CFT in $d=2$ to have deformations with mass gap all the ground states
of the Ramond sector should have equal left- and right-moving $U(1)$ charges.}
This relation can be used to derive the BPS jumping phenomenon: The reason
is that the RHS is fixed.  Thus the LHS should also remain fixed.  However, as we change
the couplings of the theory, the central charges may change and when the phases
of the central charges reorder, the fact that the monodromy does not change implies
that the BPS numbers should change so that
$$S_{ab} S_{ac}S_{bc}=S_{bc}S'_{ac}S_{ab}$$
where $S,S'$ denote the corresponding operators before and after the wall-crossing respectively.
This leads to the degeneracy changing formula given above.\footnote{This formula {was originally} derived
under the assumption that only two central charges align at the same time. The argument in the present paper does not require that assumption, since it is directly expressed in terms of the group element $S_{\phi,\phi^\prime}$, namely the
phase--ordered product of the contributions from all BPS states having phases in the interval $(\phi^\prime,\phi)$,
which will be interpreted in section 4 as a quantum evolution kernel. This is important since the situations encountered
in $\cn=2$ theories in $4d$ involve the case where infinitely many phases align at the same time.}

The basic aim of this paper is to formulate a supersymmetric path-integral which computes
$I_k$, and use the invariance of the path-integral, upon deformations, to get from the
R-charge computation in the UV to the soliton counting problem in the IR.

The form of (\ref{phase}) is very suggestive:  The phase ordered product
should arise as a time ordered product in the physical theory, as was recently
proposed in the context of $4d$ wall-crossing \cite{4dwallcrossing}.  On the other
hand the phase of the BPS states correspond, at the conformal point, to the
R-charge rotation.  Since R-charge is conserved at the conformal point, it
suggests that we compute the path integral with the insertion of 
$${\rm exp}(2\pi i kR)$$
Indeed if we compute 
$${\rm Tr}\Big[(-1)^F g\, {\rm exp}(-\beta H)\Big]$$
where $g$ is a symmetry (commuting with $H$) and which also commutes with the supersymmetry charges,
then as argued in \cite{Wittenindex} it receives contributions only from the ground states, and moreover it
does not change under compact deformations of the theory commuting with $g$.  This motivates one to consider
in our case
$$g=(-1)^{kF}  {\rm exp}(2\pi i kR)$$
The inclusion of $(-1)^{kF}$ is to make sure supercharges, which have $\pm1/2$ charge under $R$, commute
with $g$.  Indeed at the conformal point this would precisely give us $I_k$ defined above (noting that
the Ramond ground states have zero fermion number because $F=q_L-q_R$).
$$I_k={\rm Tr} \Big[(-1)^{(k+1)F}  {\rm exp}(2\pi i kR)\,  {\rm exp}(-\beta H)\Big]$$
However this definition does not quite work away from the conformal fixed points, because
$R$ is not a symmetry!  Thus our main challenge is to define an index away from the CFT
fixed points which reduces to the above definition at the fixed points.
The fact that the time ordered expression we are after is making jumps at specific
times (corresponding to phases for which BPS masses exist) in the IR suggests
that we are dealing with a time-dependent evolution operator.  In other words
it is not going to be a conventional supersymmetric theory whose Hamiltonian
is time independent.  The idea that this can be done is motivated by the simple
observation that, if we consider deformations of a conformal theory, we can always
promote the parameters to be time-dependent, thus compensating
for the violation of R-symmetry. The resulting supersymmetry turns out to be a far--reaching time--dependent generalization of the Parisi--Sourlas one \cite{parisi1,cecotti1,Parisi2,cecotti2}. 
\medskip

The organization of this paper is as follows:  In section 2 we formulate the basic setup
in the context of Landau-Ginzburg theories.  In section 3 we formulate a novel
supersymmetric system which is time-depedent.  In section 4 we evaluate the
path-integral in the UV and IR limits and recover the formula (\ref{maineq}).
In section 5 we present our conclusions.  In Appendix A we discuss some subtleties arising in the cases where many classical vacua are alligned in the $W$--plane.  In Appendix B we discuss some connections with tt* geometry.


\section{The Basic Setup}

We will be interested in 2 dimensional theories with $\cn=2$ supersymmetry with isolated vacua
and mass gap.  In such theories one can consider Hilbert spaces ${\cal H}_{ab}$ corresponding to real
line, where on the left we approach the vacuum $a$ at infinity and on the right the vacuum $b$.  The supersymmetry
algebra will in general have a central charge in this sector represented by the complex number $Z_{ab}$.  
This means that
$$E_{ab}\geq |Z_{ab}|$$
where $E_{ab}$ is the energy eigenvalue in this sector.
The BPS kinks in this sector will saturate this bound.  One can view the RG flow as acting
on the central charges by $Z_{ab}\rightarrow \lambda Z_{ab}$, where the IR limit corresponds
to $\lambda\rightarrow \infty$.
We will often assume that the theory comes from deformations of an $\cn=2$ CFT
by relevant deformations.   The conformal theory is then recovered in the limit $\lambda \rightarrow 0$.

The constructions of this paper can be done in full generality only assuming this structure.
However, for simplicity of presentation, it will be convenient to illustrate the construction
in the context of $\cn=2$ Landau-Ginzburg theories.  We will comment on the construction
in the general setup later in the paper.

LG theories have chiral superfields, which we denote by $X^i$, and the action for them involves
D-term (integrating over the full superspace) and F-terms (integrating over half the superspace).
D-term is characterized by a real function $K(X^i,{\overline X^i})$, and the F-term by
the holomorphic superpotential $W(X^i)$.   The action is given by
$$S=\int d^4\theta d^2x\, K(X^i,{\overline X^i}) +\Big[\int d^2\theta d^2x \,W(X^i)+c.c.\Big].$$
For most of our discussion we take $K=\sum_i X^i{\overline X^i}$. Our results will not depend on this assumption.
The bosonic part of the action can be written as
$$\int d^2x \Big(\partial X^i {\overline \partial X^i}+|\partial_i W|^2\Big).$$
This, in particular, implies that the vacua are given by the critical points of $W$, which we assume
to be isolated, and denote them by $X_a^i$.  Then the central charge in the $ab$ sector is given by
$$Z_{ab}=W(X_a^i)-W(X_b^i).$$
When $W$ is a quasi-homogeneous function of $X^i$, \textit{i.e.,}
$$W(\lambda^{q_i}X^i)=\lambda\, W(X^i),$$
it is expected that (with a suitable choice of $K$) this theory
corresponds to a conformal theory \cite{martinec,vw} with central charge
$$\hat c =\sum_i (1-2q_i).$$
 In this case the R-charge of the field $X^i$ is $q_i$.
 The chiral ring of this theory is identified with
$${\cal R}= \mathbb{C}[X^i]/\{ dW=0 \}, $$
and its elements are in one to one correspondence with the ground states
of the Ramond sector (whose R-charges are shifted by ${\hat c}/2$).  In particular
this leads to
$$Tr_{Ramond\ Ground \ State} t^{R}=t^{-\hat c/2}\prod_i {(1-t^{1-q_i})\over (1-t^{q_i})}.$$
In this context we can compute the supersymmetric index
$$I_k={\rm Tr} \Big[(-1)^{(k+1)F}  {\rm exp}(2\pi i kR)\,  {\rm exp}(-\beta H)\Big]$$
and find
$$I_k=\prod_{kq_i\in {\bf Z}} {(1-q_i)\over q_i}.$$
The formulation of $I_k$ as the dimension of a cohomological problem at the conformal fixed point, and
its relation to LG orbifolds was studied in \cite{lgorbifold}.

In general we will assume that $W$ is a deformation
away from quasi-homogeneous form by relevant deformations, giving rise to isolated vacua.
Rescaling $W\rightarrow \lambda W$ corresponds to the RG flow in this context, and the conformal
point, where $W$ is quasi-homogeneous is the fixed point of this transformation.

In defining the relevant supersymmetric theory, we will be compactifying this theory first on a circle.
It turns out that the radius of the circle does not enter the computation, and we can consider
the case of the theory reduced to 1-dimension, \textit{i.e.}\! the quantum mechanics theory with 4
supercharges.  Later in the paper we show why our construction does not change when we
go to the full 2d theory.

\section{A Time-Dependent SQM:  Formulation of an Index}

\subsection{The index $I_k$}

We consider the reduction of the LG theory to 1d.  We first consider the quasi-homogeneous case, where we have
an $\cn=2$ SQM system (4 supercharges) with an $R$--symmetry with generator $R$, under which the supercharges have charges $\pm \frac{1}{2}$. 
To make this section self-contained,
let us review the construction of the index $I_k$ in this context:  Let $k$ be an integer, and consider
\begin{equation}\label{indexIk}
	I_k= \mathrm{Tr}\Big[ e^{2\pi i\, k\, R}\, (-1)^{(k-1)F}\, e^{-\beta H}\Big].
\end{equation}
This expression is an index receiving contributions only from the supersymmetric vacua. Indeed, non--zero energy states appear in multiplets of $R$--charges $(r, r+\tfrac{1}{2},r-\tfrac{1}{2}, r)$ and the factor $e^{2\pi i\, k\, R}\, (-1)^{(k-1)F}$ is equal to
\begin{equation}
e^{2\pi i\, k\, R}\, (-1)^{(k-1)F}=	
\begin{cases}
\pm e^{2\pi i k r} & \text{bosonic states}\\	
\mp e^{2\pi i k r} & \text{fermionic states}\\
\end{cases}
\end{equation}
so that the total contribution of each non--trivial supermultiplet is zero. (The same conclusion remains true even in presence of central charges in the supersymmetric algebra).

Then
\begin{equation}
	I_k = \sum_\mathrm{vacua} (-1)^{(k-1)F}\, e^{2\pi i k\, r}.
\end{equation}
\smallskip

As in the last section we consider a LG model with superpotential $W(X_i)$ which is quasi--homogeneous in the chiral superfields $X_i$ with weights
$q_i$, that is, $W(\lambda^{q_i}X_i)=\lambda\, W(X_i)$, with $q_i>0$. 
\medskip

The index $I_k$ may be represented by a path--integral with the boundary conditions
\begin{gather}
 X_i(\beta)= e^{2\pi i k q_i}\, X_i(0)\\
\psi_i(\beta)=e^{2\pi i k q_i}\, \psi_i(0),
\end{gather}
under which the superpotential is periodic
\begin{equation}
 W(X_i(\beta))= W(e^{2\pi i k q_i}X_i(0))= e^{2\pi i k} W(X_i(0))= W(X_i(0)).
\end{equation} 
Using the periodicity of $W$, we may rewrite the action in the form
\begin{equation}\label{actionuseform}
 \int\limits_0^\beta dt\Big(\dot X_i + e^{-i\alpha}\, \overline{\partial_i W}\Big)\Big(\dot X_i^*+ e^{i\alpha}
\partial_i W\Big) +\text{fermions},
\end{equation}
since the difference with the usual action is
\begin{equation}
 2\, \mathrm{Re}[e^{i\alpha}(W(\beta)-W(0))]=0.
\end{equation} 

We may also rewrite the fields as
\begin{gather}
X_i(t)= e^{2\pi i k q_i t/\beta}\, Y_i(t)\\ 
\psi_i(t)=e^{2\pi i k q_i t/\beta}\,\chi_i(t)
\end{gather}
 where $Y_i(t)$ and $\chi_i(t)$ are strictly periodic. In the new variables, the bosonic action becomes
\begin{equation}
 \int\limits_0^\beta \sum_i\Big|\dot Y_i + \tfrac{2\pi i k q_i}{\beta}Y_i+ e^{-i(\alpha+2\pi k t/\beta)}\, \overline{\partial_i W(Y_i)}\Big|^2\: dt
\label{boson}
\end{equation}
and the fermionic part
\begin{equation}\begin{split}
 \int\limits_0^\beta dt \Big(&\tilde\chi_{\bar i}(d_t \chi_i+\tfrac{2\pi i k q_i}{\beta}\chi_i)+
\tilde\chi_{i}(d_t \chi_{\bar i}-\tfrac{2\pi i k q_i}{\beta}\chi_{\bar i})+\\
&+
e^{i(\alpha+2\pi k t/\beta)} (\partial_i\partial_j W(Y))\tilde\chi_{\bar i}\chi_{\bar j}+
e^{-i(\alpha+2\pi k t/\beta)} (\partial_{\bar i}\partial_{\bar j} \overline{W}(\bar Y))\tilde\chi_{i}\chi_{j}\Big).
\end{split}
\label{fermion}
\end{equation}

Even though we were motivated to formulate the above path-integral starting from the conformal case,
where $W$ is quasi-homogeneous, {\it the path-integral we have ended up in this formulation
makes sense even when} $W$ {\it is deformed by relevant terms away from the quasi-homogeneous limit}.
We thus consider the above action for arbitrary deformed $W$\footnote{Rewriting this
action in terms of $X^i, \psi^i$ gives the usual LG Lagrangian with two differences:  The
superpotential
$W$ is now time-dependent, and we have an additional term given 
 by $[-\int dt\, e^{i \alpha}{\partial W\over \partial t} +c.c.]$.}.

The resulting path integral is an invariant \textit{index}. This follows from the fact that it has a supersymmetry, albeit not of the standard kind\footnote{\ The system is not invariant under time translations, so the square of  a supersymmetric transformation is not a translation in time.}. Rather, our path integral is invariant under a generalized version of the Parisi--Sourlas supersymmetry \cite{parisi1,cecotti1,Parisi2,cecotti2}.   The extension of the Parisi-Sourlas techniques
to a {\it  time-dependent} situation, which is what we need for the present work, is novel.
Let us briefly review how Parisi-Sourlas supersymmetry works in the usual setting. 
\bigskip

\subsection{Parisi--Sourlas supersymmetry}
Writing
\begin{equation}\label{stoc}
 h_i =\dot Y_i + \tfrac{2\pi i k q_i}{\beta}Y_i+ e^{-i(\alpha+2\pi k t/\beta)}\, \overline{\partial_i W(Y_i)} 
\end{equation} 
the bosonic action \eqref{boson} becomes simply
\begin{equation}\label{bosaction}
 S_B=\int\limits_0^\beta \sum_i |h_i|^2\,dt,
\end{equation} 
while the fermionic action is
\begin{equation}\label{fermionicaction}
 S_F=\int\limits_0^\beta \begin{pmatrix} \tilde\chi_j & \tilde\chi_{\bar j}\end{pmatrix}
\begin{pmatrix} \frac{\delta h_{\bar j}}{\delta Y_{i}} & \frac{\delta h_{\bar j}}{\delta Y_{\bar i}}\\
\frac{\delta h_{j}}{\delta Y_{i}} & \frac{\delta h_{j}}{\delta Y_{\bar i}}\\ \end{pmatrix}
 \begin{pmatrix} \chi_i \\ \chi_{\bar i}\end{pmatrix}\: dt.
\end{equation} 

A Parisi--Sourlas supersymmetric system is defined by an action of the form
\begin{equation}
	S=S_B+S_F
\end{equation}
where $h_i=h_i[Y_j]$ is any functional map from the original bosonic fields $Y_i$ to the Gaussian fields $h_i$
(this map is also known as the Nicolai map \cite{nicolai,cecotti1}).\smallskip

By construction, a general Parisi--Sourlas system with Lagrangian\footnote{ For generality, we write the Lagrangian in the real notation corresponding to a general $\mathcal{N}=1$ SQM model. If the underlying model has $\mathcal{N}=2$ \textsc{susy}
everything gets complexified as in eqns.\eqref{bosaction}\eqref{fermionicaction}.}
\begin{equation}\label{PSlagrangian}
	L= \frac{1}{2}h_i\,h_i+\bar\chi_i\, \frac{\delta h_j}{\delta Y_i}\,\chi_j,
\end{equation}
is invariant under the tautological supersymmetry
\begin{align}
	\delta Y_i &=\bar \chi_i\,\epsilon\\
	\delta \chi_i &=- h_i\,\epsilon\\
	\delta\bar \chi_i& =0,
\end{align}
 where $\epsilon$ is a complex Grassmannian parameter.  Here we have written the
action in the case of real fields --- our case can be recovered from this by viewing the index of the
fields to also label the complex conjugate fields (see \S  3.3).
In the usual time--independent setting the Lagrangian \eqref{PSlagrangian} is actually Hermitian; in that case the model is invariant under a second supersymmetry, namely the Hermitian conjugate of the above one. For special forms of the functionals $h_i[Y]$ the supersymmetry enhances further.
 \smallskip
 
In any Parisi--Sourlas system \cite{parisi1,cecotti1,Parisi2,cecotti2}, the Gaussian integral over the fermions produces precisely the Jacobian determinant\footnote{Up to a crucial sign. See discussion below.} for the functional change of variables 
$Y_i\rightarrow h_i$, and then the full path integral takes in the new fields the Gaussian form
\begin{equation}\label{gaussianintegral}
 \int [dh_i] \exp\Big[-\frac{1}{2}\int_0^\beta \sum_i h_i^2\, dt\Big].
\end{equation} 
Thus the $h_i(t)$ are Gaussian fields with exact correlation functions
\begin{align}
	&\langle h_i(t)\, h_j(t^\prime)\rangle =\delta_{ij}\: \delta(t-t^\prime),
\end{align}
also known as ``white noise'' fields.
Then the Nicolai map $h_i[Y_j]=h_i$ is interpreted as a stochastic differential equation with a white noise source. For instance, the map in  eqn.\eqref{stoc} with $k=0$ (the corresponding index, $I_0$, being the usual Witten index) is the standard Langevin equation
for a `drift potential' equal to $2\,\mathrm{Re}(e^{i\alpha}W)$. In appendix B we shall relate the Fokker--Planck equation associated to such a Langevin equation to the tt* geometry of the corresponding $\cn=2$ model. 
\smallskip

In $1d$ all supersymmetric LG models are, in particular, Parisi--Sourlas systems. The same is true in $2d$, provided the LG model has $\cn=2$ supersymmetry \cite{Parisi2,cecotti2}. However, in these dimensions, there are many Parisi--Sourlas supersymmetric systems which are not equivalent to the standard ones. The models relevant for the present paper are a special instance of such more general supersymmetric systems.\smallskip 

In a Parisi--Sourlas supersymmetric model, the Witten index $\Delta$ is interpreted \cite{cecotti1} as \textit{the degree of the stochastic map}
\begin{equation}\label{stomap}
	Y_i\rightarrow h_i\equiv h_i[Y_j],
\end{equation}
that is, $\Delta$ is the number of the periodic solutions $Y_i(t)$ of the equation
\begin{equation*}
	h_i[Y_i(t)]=h_i(t),
\end{equation*}
for a given \textit{generic} periodic function $h_i(t)$.
Indeed, $\Delta=\mathrm{Tr}[(-1)^F \exp(-\beta H)]$ is equal to the path--integral with all fields periodic, and hence, after the integration of the fermions, it is given by the Gaussian integral \eqref{gaussianintegral}. Since the Gaussian integral (computed over the space of the periodic functions) is $1$, the value of the original path integral is given just by the \textit{net} number of times that the space of periodic paths $Y_i(t)$ covers, under the map $Y_i(t)\rightarrow h_i(t)$, the space of the periodic functions in the functional $h_i$--space \cite{cecotti1}, which is what is meant by the degree of the functional map.
Mathematically what this means is that we have a map from the infinite dimensional space of loops in $\mathbb{C}^N$
to itself, given by $h$, and we simply compute the degree of this map\footnote{\ In general there may be subtleties \cite{cecotti3} in defining the degree of this infinite dimensional map.   However, if the `drift prepotential' $W$ is holomorphic, or if our system is a compact deformation of such a model, we
do not have to worry about such pathologies.} \eqref{stomap}.

The degree, being the analog of the Witten index for the present theory, is invariant under the continuous deformations of the superpotential which do not change its leading behavior at infinity.
\smallskip

\subsection{Time-dependent supersymmetry}\label{td}

From the above discussion, it is obvious that the path integral of a Parisi--Sourlas system is supersymmetric even if the functional $h_i[Y_j]$ has an explicit time--dependence, provided this dependence is periodic with the same period $\beta$ appearing in the boundary conditions for the bosonic/fermionic fields. This applies, in particular, to our model, defined by eqns.\eqref{stoc}\eqref{bosaction}\eqref{fermionicaction} where the fields $Y_i(t)$, $\chi_j(t)$ are now strictly periodic of period $\beta$.  

Explicitly, the action $S=S_B+S_F$ is invariant under the supersymmetry
\begin{align}
 \delta Y_i & = \chi_i\, \epsilon & \delta \overline{Y}_{\bar i}&= \chi_{\bar i}\, \epsilon\label{tdsusy1}\\
\delta \tilde\chi_j&= h_j\, \epsilon & \delta \tilde\chi_{\bar j}& =h^*_{\bar j}\, \epsilon\\
\delta \chi_j&=0 & \delta\chi_{\bar j}&=0.\label{tdsusy3}
\end{align}
where $h_j$ (resp.\! $h^*_{\bar j}$) stands for the \textsc{rhs} of eqn.\eqref{stoc} (resp.\! its complex conjugate).
\smallskip

We are thus led to study this novel time-dependent supersymmetric quantum mechanical system.\smallskip

Then our index $I_k$ is the Witten index for the generalized Parisi--Sourlas system defined by the modified Langevin equation \eqref{stoc} or, equivalently, the degree of this stochastic map. In particular, this implies that $I_k$ is an \emph{integer,} a property that is not immediately obvious from its definition \eqref{indexIk}.

\subsection{The underlying time--dependent TFT}

In this paper we have adopted the language of the `physical' theory. However many of the manipulations in section 4 below could have been done equally well in the framework of the (time--dependent)
`topological' QM theory \cite{gozzi}. The topological version of the Parisi--Sourlas model \eqref{PSlagrangian} is given by the Lagrangian
\begin{equation}\label{topLag}
	L= F_i\,h_i[Y]+\bar\chi_i\, \frac{\delta h_j}{\delta Y_i}\,\chi_j
\end{equation}
where now $F_i$ is an independent (auxiliary) field\footnote{Of course, the auxiliary fields $F_i$ may be introduced also in the `physical' theory to make the supersymmetry in eqns.\eqref{tdsusy1}--\eqref{tdsusy3} to be nilpotent off--shell.}.
The Lagrangian \eqref{topLag} is invariant under the nilpotent topological supersymmetry
\begin{align}
	\delta Y_i&= \bar\chi_i\epsilon & \delta F_i &=0\\
	\delta \chi_i&=-F_i \epsilon & \delta\bar\chi_i&=0.
\end{align}
The functional integral over the auxiliary field $F_i$ produces a delta--function enforcing the differential equation
$h_i[Y]=0$. That is, in the topological theory the stochastic Langevin equation gets replaced by the corresponding deterministic differential equation obtained by setting the ``noise'' to zero. Indeed, in the old days \cite{gozzi} this topological version was seen as a trick to give a path integral representation to the solutions of classical deterministic systems.\medskip

Let $\co[Y]$ be an observable depending on the bosonic fields $Y_i$. Integrating away the fermions and the auxiliary field
\begin{multline}
 \big\langle\, \co[Y]\,\big\rangle_\mathrm{topologic}= \int [dY_j] \det\left[\frac{\delta h_i}{\delta Y_j}\right]\: \delta\Big[ h_i[Y]\Big]\:\co[Y]=\\
=\int [dh_i]\,\frac{\det\left[\frac{\delta h_i}{\delta Y_j}\right]}{\Big|\det\left[\frac{\delta h_i}{\delta Y_j}\right]\Big|}\: \delta[h_i]\: \co\big[Y[h]\big]= \sum_{\mathrm{classical}\atop \mathrm{solutions}}\pm\, \co[Y_\mathrm{clas.}].\qquad
\end{multline}  

In particular, the path--integral with periodic boundary conditions gives
\begin{multline}
 \int\limits_\mathrm{periodic} [dY_j] \det\left[\frac{\delta h_i}{\delta Y_j}\right]=\\
= 
\text{\textit{net} }\#\!\begin{pmatrix}\text{classical periodic solutions of}\\ \text{the differential equation } h_i=0\end{pmatrix}\equiv \Delta,\qquad
\end{multline} 
where by the \textit{net} number of solutions we mean the difference between the number of positively and negatively oriented solutions.  More generally, the transition amplitudes
\begin{equation}
 \big\langle\, Y_{(f)}, t^\prime\,|\, Y_{(i)}, t\, \big\rangle_\mathrm{topological} =\int\limits_{Y_j(t)=Y_{(i)}}^{Y_j(t^\prime)=Y_{(f)}} [dY_j] \det\left[\frac{\delta h_i}{\delta Y_j}\right]
\end{equation} 
are given by the \textit{net} number of classical solutions to the differential equations $h_i[Y]=0$ satisfying the corresponding boundary conditions.
Some of the manipulations of section 4 may be easily rephrased in this topological language.

\section{Evaluation of the Index}

In the previous section we formulated a path--integral representation of the index $I_k$.
As discussed, its value does not depend on small deformations of the superpotential. In particular,
varying the parameters which flow the theory from UV to the IR limit is such a deformation.

\subsection{UV limit}

To study the UV limit, we decompose $W(Y_i)$ as
\begin{equation}
 \lambda W(Y_i)= \lambda \Big(W_1(Y_i)+ \sum_{q<1} W_q(Y_i)\Big),
\end{equation} 
where the term $W_q(Y_i)$ is quasi--homogeneous of weight $q$. The deformation is `relevant' and hence does not change the behaviour at $\infty$ and the degree of the Nicolai map. Under the redefinition $Y_i\rightarrow \lambda^{q_i} Y_i$, the
superpotential becomes
\begin{equation}
 W_1(Y_i)+\sum_{q<1} \lambda^{1-q}\, W_q(Y_i).
\end{equation} 
Therefore, as $\lambda\rightarrow 0$, the theory becomes conformal and we get the result for the quasi--homogeneous superpotential $W_1(Y_i)$, and hence the index
\begin{equation*}
I_k=\sum_\mathrm{vacua} (-1)^{(k-1)f}\,e^{2\pi i k r}.	
\end{equation*}

\subsection{IR limit: path--integral analysis}

The limit $\lambda\rightarrow\infty$ corresponds to the IR limit and
should give the same answer for the index. The bosonic action is rewritten as
\begin{equation}
 \int\limits_0^\beta \lambda^2\, \Big|\frac{1}{\lambda}\frac{d Y_i}{dt} + \tfrac{2\pi i k q_i}{\beta\lambda}Y_i+ e^{-i(\alpha+2\pi k t/\beta)}\, \overline{\partial_i W(Y_i)}\Big|^2\, dt,
\end{equation} 
and, in the limit $\lambda\rightarrow \infty$, the path--integral should be saturated by the configurations satisfying  
\begin{equation}\label{eqmotionbps}
 \frac{1}{\lambda}\frac{d Y_i}{dt} + \tfrac{2\pi i k q_i}{\beta\lambda}Y_i+ e^{-i(\alpha+2\pi k t/\beta)}\, \overline{\partial_i W(Y_i)}=0.
\end{equation} 

Let us consider the (Euclidean) transition amplitudes
\begin{equation}\label{amplitude}
 \langle Y_i^\prime, t_0+t\:|\: Y_i, t_0-t\rangle = \int\limits_{Y_i\: t_0+t}^{Y^\prime_i,\: t_0-t} [d\text{(fields)}]\, e^{-S_E}
\end{equation} 
from a configuration $Y_i$ at time $t_0-t$ to a configuration $Y^\prime_i$ at $t_0+t$ which are defined by the above path integral
(of course, these are \emph{not} the physical amplitudes for the original quantum system).\smallskip 

As $\lambda\rightarrow \infty$
the \textsc{rhs} of eqn.\eqref{amplitude} gets saturated by the solutions to eqn.(\ref{eqmotionbps}) having the right boundary conditions. We set $\tau= 2\pi k\lambda(t-t_0)/\beta$,
$\hat \alpha= \alpha+2\pi k t_0/\beta$, and $\mu= \beta/(2\pi k)$. Then 
\begin{equation}\label{BPSeqfin}
 \frac{dY_i}{d\tau}+i\frac{q_i}{\lambda}Y_i+\mu\, e^{-i(\tau/\lambda+\hat \alpha)}\,\overline{\partial_i W(Y_i)}=0. 
\end{equation}   
As $\lambda\rightarrow\infty$, this equation becomes the one describing a BPS soliton of phase $e^{i\hat \alpha}$.
Writing $Y_i(\tau)=e^{-i q_i\tau/\lambda}X_i(\tau)$, 
\begin{equation}
 \frac{dX_i}{d\tau}+ \mu e^{-i\hat \alpha}\sum_q e^{i(q-1)\tau/\lambda} \,\:\overline{\partial_i W_q(X_i)}=0.
\end{equation} 
One looks for an asymptotic solution as $\lambda\rightarrow\infty$. If $t$ in the \textsc{lhs} of (\ref{amplitude}) is much smaller than $\beta$, so that $\tau/\lambda\ll 1$ everywhere along the paths contributing to the path integral in the \textsc{rhs} of eqn.(\ref{amplitude}), the $O(\lambda^{-1})$ corrections to the BPS soliton equation \eqref{BPSeqfin} remain small as 
the fields go from one vacuum to the other. Measured in units of $\tau$, the small time $2t$ becomes of order $O(\lambda)$, and hence infinitely long as $\lambda\rightarrow\infty$, so there is plenty of $\tau$ time to complete the transition from one asymptotic vacuum to the other one interpolated by the given BPS soliton. 
Indeed, there is plenty of rescaled time to accomodate a \textit{chain} of BPS kinks, corresponding to multiple jumps from one vacuum to the next one, provided all the involved solitons have the (same) appropriate BPS phase. 
The saturating configuration differs from a classical vacuum by
a quantity of order $O(e^{-M\tau})=O(e^{-C \lambda (t-t_0)})$. Therefore, as $\lambda\rightarrow\infty$ the saturating configuration is a vacuum, \textit{except} for a time region of size $O(1/\lambda)$ (measured in `physical' time $t$) around the special times $t_0$ at which $e^{i\hat\alpha}=e^{i\alpha}\, e^{2\pi i k t_0/\beta}$ is the phase of a BPS soliton.
\smallskip

In conclusion, in the limit $\lambda\rightarrow\infty$, for {almost all times $t$, $t^\prime$, with  $t-t^\prime\ll \beta$},
one may effectively replace the quantum amplitude (\ref{amplitude}) with a finite
$n\times n$ matrix (where $n$ is the number of supersymmetric vacua)
\begin{equation}\label{amplsmalltime}
 \langle Y_i(a), t^\prime|Y_i(b), t\rangle\rightarrow (S_{t^\prime, t})^a_{\ b}= {\delta^a}_b+ (N_{t^\prime,t})^a_{\ b}
\end{equation}  
where the \emph{integer} $(N_{t^\prime,t})^a_{\ b}$ counts \emph{with signs} the number of different BPS kink \textit{chains} connecting vacuum $b$ to vacuum $a$ though a sequence of intermediate vacua
of the form
\begin{equation}\label{chain3}
	a \equiv a_0 \rightarrow a_1\rightarrow a_2 \rightarrow \cdots \rightarrow a_m \equiv b,
\end{equation}
which satisfy the condition of well--ordered central charge phases\footnote{The symbol $\leq$ stands for the natural order of increasing phases.} 
\begin{multline}\label{cenphasecond}
	e^{i\alpha}\, e^{2\pi i k t/\beta} \leq \mathrm{phase}(\overline{W}_{a_1}-\overline{W}_a) \leq
\mathrm{phase}(\overline{W}_{2}-\overline{W}_1) \leq\cdots\\
\cdots \leq \mathrm{phase}(\overline{W}_{b}-\overline{W}_{m-1}) \leq
e^{i\alpha}\, e^{2\pi i k t^\prime/\beta}, \qquad
\end{multline}
where each vacuum transition $a_i\rightarrow a_{i+1}$ is triggered by a BPS state of the appropriate phase.
\smallskip

In the special case that the time interval $t^\prime-t$ is small enough that the angular sector
\begin{equation}\label{ttarc}
 \big(e^{i\alpha}\,e^{2\pi i k t/\beta},\: e^{i\alpha}\,e^{2\pi i k t^\prime/\beta}\big)
\end{equation}
contains only one BPS phase $\frac{1}{2}\log(\overline{Z}_{ba}/Z_{ba})$, and this phase is associated to a unique pair\footnote{This is the generic situation for $2d$ LG models.} of vacua $b\,a$, the integer $(N_{t^\prime,t})^a_{\ b}$ is just \emph{plus or minus} the number of BPS solitons connecting the two vacua $a$, $b$, thus reproducing eqn.\eqref{SINT}. However, our present discussion, in terms of the integral--valued time--evolution kernel $S_{t^\prime,t}$, is more general than eqn.\eqref{SINT} and holds for any number of BPS rays and also for vacua which are aligned in the $W$--plane.\smallskip

The ambiguity in the sign of $(N_{t^\prime,t})^a_{\ b}$ reflects the relative signs of the bosonic and fermionic determinants around a given saturating configuration. Supersymmetry guarantees that these two determinants are equal in \textit{absolute value}, but not necessarily in sign\footnote{The bosonic determinant is the square root of the square of the fermionic one, so it is always positive, while the fermionic one may, in principle, be negative.} (the path integral is essentially a Witten index which must be an integer, but may be a \textit{negative} integer). We shall formalize the correct \emph{sign rule} for $(N_{t^\prime,t})^a_{\ b}$ momentarily.\medskip

The time--evolution kernels satisfy a group--law, namely
\begin{equation}\label{grouplaw}
 (S_{t^\prime, t})^a_{\ b}= (S_{t^\prime, t^{\prime\prime}})^a_{\ c}\,(S_{t^{\prime\prime},t})^c_{\ b}\qquad \text{for }t^{\prime\prime}\in (t, t^\prime).
\end{equation}
Then, even if our IR computation was reliable for time intervals $|t^\prime-t|\ll \beta$, the result can be extended to arbitrary long time--intervals by defining
\begin{equation}\begin{split}
 &S_{t_m, t_0}= S_{t_m, t_{m-1}}\, S_{t_{m-1},t_{m-2}}\,\cdots\, S_{1,t_0}\\
&\text{where}\quad 0<t_1<t_2<\cdots < t_m, \quad\ t_{i+1}-t_i\ll\beta. 
\end{split}\end{equation} 
 The discussion between eqn.\eqref{amplsmalltime} and eqn.\eqref{cenphasecond} applies to arbitrary time intervals. In particular, (as $\lambda\rightarrow\infty$) all amplitudes are given by integer numbers.
\medskip

$(N_{t^\prime,t})^a_{\ b}$ is a strictly \textit{triangular} matrix (in a suitable basis) provided the time interval $(t,t^\prime)$ satisfies
\begin{equation*}
	0< (t^\prime-t)<\frac{\beta}{2 k}.
\end{equation*}
Indeed, if the phase of $\overline{Z}_{ba}$ belongs to the corresponding angular sector \eqref{ttarc},
 so that $(N_{t^\prime,t})^a_{\ b}$ is possibly non--zero, the phase of $\overline{Z}_{ab}=-\overline{Z}_{ba}$ 
belongs to the complementary angular sector, and then $(N_{t^\prime,t})^b_{\ a}=0$.
\medskip

Consider the amplitude translated in time by $\beta/2k$, 
\begin{displaymath}
 S_{t^\prime+\beta/2k,\, t+\beta/2k}.
\end{displaymath}
It counts BPS states with phases in the \textit{opposite} angular sector with respect to the one contributing to $S_{t^\prime, t}$, that is, it counts the states in the conjugate sectors. The \emph{absolute} number of solitons in the $ab$ sector is the same as in the sector $ba$, $|N_{ab}|=|N_{ba}|$. However, their \textit{signs} may be different. We want to argue that, indeed, they have opposite signs. For simplicity, we focus on the generic situation in which there is no vacuum alignement, and all amplitudes $S_{t^\prime,\, t}$ may be written as a time--order product of elementary ones of the form in eqn.\eqref{SINT}
\begin{displaymath}
 S_{t^\prime,\, t}= T\big(\prod S_{(i)}\big)\ \in SL(n,\mathbb{Z}),
\end{displaymath}
where $S_{(i)}=1+N_{(i)}$ with $N_{(i)}$ a nilpotent matrix of rank $1$. If the signs of the $N_{(i)}$'s of conjugate sectors were the same, a translation in time by $\beta/2k$  will act on the elementary amplitudes by simply transposing them. Under deformation to a non--generic configuration, this would correspond to the rule
\begin{displaymath}
S_{t^\prime+\beta/2k,\,t+\beta/2k}=S_{t^\prime,\, t}^T. 
\end{displaymath}
But this equation is inconsistent with the group law, since transposition is not an automorphism of $SL(n,\mathbb{Z})$. What \textit{is} a group automorphism is transposition and inverse combined, $S\mapsto (S^T)^{-1}$ (the Cartan involution),
which acts on the elementary factors as
\begin{equation}\label{signsrule2}
S_{(i)}= 1+N_{(i)}\mapsto 1-(N_{(i)})^T \equiv (S_{(i)}^{-1})^T,	
\end{equation}
that is, invertes the sign of the elementary soliton multiplicities, $N_{ab}=-N_{ba}$. Then, in full generality,
\begin{equation}\label{pctforampl}
	S_{t^\prime+\beta/2k,\, t+\beta/2k}= \big(S_{t^\prime,t\,}^{-1})^T.
\end{equation}
This equality fixes the rule for the signs (up to convention--dependent choices). A different argument for the sign rule will be presented in appendix \ref{secschoeredinger}.
\medskip

Let $S=S_{\beta/2k, 0}$ be the amplitude for the largest time interval such that it is still a triangular matrix. From eqn.\eqref{pctforampl} we have
\begin{equation}
	S_{\beta,0}=(S^{-1})^T S (S^{-1})^T S\cdots (S^{-1})^T S= \big((S^{-1})^T S)^k,
\end{equation}
and then 
\begin{equation}
 I_k =\mathrm{tr}\big[\big((S^T)^{-1}\,S\big)^k\big].
\end{equation} 
This is, of course, the well--known formula \cite{classification}.

\subsection{Extensions to Two--dimensions and More General $\cn=2$ Theories}
We have mainly focused on the one dimensional formulation of the index.  But nothing essential
changes in going to the main case of interest, which is $d=2$. One quantizes the system in a rectangular torus of sizes $L$, $\beta$ with the boundary conditions
\begin{equation}
 X_i(L,t)=X_i(0,t),\qquad X_i(x,\beta)= e^{2\pi i k q_i}\, X_i(x,0).
\end{equation} 

The $\cn=2$ $2d$ Landau--Ginzburg models may still be written as Parisi--Sourlas supersymmetric systems \cite{Parisi2,cecotti2}.
Eqn.(\ref{actionuseform}) is replaced by ($z=x+it$)
\begin{equation}\label{2dactionPS}
 \int\limits_\mathrm{torus} d^2z\,\Big(\partial_z X_i + e^{-i\alpha}\, \overline{\partial_i W}\Big)\Big({\partial}_{\overline{z}} X_i^*+ e^{i\alpha}
\partial_i W\Big) +\text{fermions},
\end{equation}
which again is equivalent, {in the quasi--homogeneous case}, to the standard action. Indeed, the action \eqref{2dactionPS} differs from the usual one by
\begin{multline}
	\frac{i}{2}\int\limits_\mathrm{rectangle}\Big( \partial(e^{i\alpha}W) \wedge d\overline{z}-\overline{\partial}(e^{-i\alpha}\overline{W})\wedge dz\Big)=\\
	=\frac{i}{2}\int\limits_{\partial(\mathrm{rectangle})}\Big(e^{i\alpha}W\, d\overline{z}- e^{-i\alpha}\overline{W}\,dz\Big)=0,
\end{multline}
since $W$ is still periodic: $W(x,0)=W(x,\beta)$ and $W(0,t)=W(L,t)$.\smallskip

Again, one can introduce strictly periodic fields $Y_i$, $\chi_i$, getting a time--dependent superpotential 
\begin{displaymath}
W(Y_i,t)= e^{2\pi i k t/\beta}\: W(Y_i). 
\end{displaymath}

$W(Y_i,t)$ is then deformed away from the quasi--homogeneous case, by adding relevant operators to $W(Y_i)$. One ends up with the generalized Parisi--Sourlas supersymmetric system defined by the stochastic partial differential equations 
\begin{equation}
	\partial_z Y_i +\frac{\pi i k}{\beta}Y_i+\lambda\, e^{-i\alpha}\, e^{2\pi k (\overline{z}-z)/\beta}\:\overline{\partial_i W}
=h_i(z,\bar z),
\end{equation}
to which the discussion in section 3 applies word for word. Again, in the limit $\lambda\rightarrow\infty$, the amplitudes are given by a time ordered product of integral matrices whose elementary entries are the multiplicities of the BPS particles with the given phase. In fact, the matrices are numerically the same as in the corresponding $1d$ models\footnote{This
construction can be extended to the case of $\cn =1$ supersymmetric theories in 4d.
These theories can have BPS domain walls, and the considerations
in this paper relate their spectrum to the R-charges at the UV fixed points.}.
\medskip

Also one can formulate the action in a more general setup than the LG case.  All we needed
for the definition of the action was to introduce a connection which couples to the R-charges at the conformal
$A_0^R=2\pi k/\beta$.
Moreover, we replace the $d^2\theta \rightarrow d^2\theta\, e^{2\pi i kt/\beta}$.  In this way we can formulate
the supersymmetric path-integral.  Furthermore one can easily show that the UV and IR computations
yield the R-charges of the ground states in the UV, and the corresponding solitons in the IR.  We thus have a general
proof of the main result for general massive deformations of $\cn=2$ theories in 2 dimensions.
\medskip

From another point of view, what we have done is to construct a quantum (time--dependent) system whose Schroedinger equation corresponds to the Lax equations whose compatibility conditions are the tt* equations for the given $\cn=2$ model; all of these structures exist for any $\cn=2$ theory and not only in the LG case.

\section{Conclusion}

In this paper we have seen how one can twist the path-integral of a supersymmetric theory
with R-symmetry.  It was known that this object exists at the conformal point.  The novelty here is that we have found a way to extend R-twisting 
of the path-integral  even after we deform the action by relevant terms that break R-symmetry.   As a by-product
we have found a simple picture of how the 2d wall-crossing in the $\cn=2$ theories work, and why it is related to R-charges
of the corresponding conformal theory.  

The ideas in this paper should apply to many other cases.  For example for $d=4$ theory with
$\cn=2$ supersymmetry, a similar picture should hold, which gives another derivation
of Kontsevich-Soibelman wall-crossing result \cite{KS}, very much in line with the string theoretic
derivation proposed in \cite{4dwallcrossing}.  Indeed similar ideas seem to hold there which relates
the $d=4$ wall-crossing to the 2d case \cite{cnv}.

One could also consider other CFT's  where R-symmetry is non-abelian, say $SU(2)$.  In such
a case one can consider more interesting twistings along a higher dimensional
geometry than a circle.  It would be very interesting to study potential implications of such twists
for conformal field theories and their deformations.

\section*{Acknowledgements}

We would like to thank Andy Neitzke for valuable discussions.  The research of
CV was supported in part by NSF grant PHY-0244821.

\appendix
\section{The case of collinear vacua}

The arguments in the main body of the paper apply even in models having a plurality of distinct solitonic sectors $ab$ with the same phase of the central charge $Z_{ab}$. These (non generic) models have some particular property, which are universal for a given vacuum alignment geometry in the complex $Z$--plane. In this appendix we discuss these properties in the light of the arguments of the present paper. The conclusions are confirmed by the analyis of the explicit models whose tt* equations can be solved in closed form.\medskip

\subsection{The $1/n$ rule}\label{1nrule}
\smallskip 

The case of three aligned vacua was already discussed in appendix B of \cite{classification}. There it is shown that, if we have
three vacua with critical values $W_1, W_2, W_3$ lying (in this order) on a straight line in the $W$--plane, and there is one physical BPS soliton connecting vacua $1,2$
and one connecting vacua $2,3$, but no soliton connecting the extremal vacua $1,3$, then the leading IR behaviour of the index
\begin{equation*}
 Q_{13}=\mathrm{Tr}_{13}\Big[(-1)^F\, F\, e^{-\beta H}\Big],
\end{equation*} 
would look like as we had $\pm \frac{1}{2}$ worth of BPS solitons of mass $|Z_{13}|$. At first this looks puzzling, since the number of particles should be integral. However, this strange result is actually \textit{required} by the integrality of the spectrum. As shown in the present paper, what actually must be integral are the group elements $S_{t^\prime, t}\in SL(n,\mathbb{Z})$. As stressed by Kontsevich and Soibelman \cite{KS}, and as it is manifest from our eqn.\eqref{finschoerdinger}, the `soliton multiplicities' $\mu_{ab}$, as defined by the IR asymptotics of the index $Q_{ab}$, belong to the corresponding Lie algebra $\mathfrak{sl}(n,\mathbb{Q})$. Of course,
we pass from the Lie algebra to the group by taking the exponential; we shall write
\begin{equation}
 S_{t^\prime,t}=\exp\!\big(L_{t^\prime,t}\big),
\end{equation}
where $L_{t^\prime,t}\in \mathfrak{sl}(n,\mathbb{Q})$ represents the $Q$--index `multiplicities' $\mu$ for the $ab$ sectors having phases in the angular sector associated to the time interval $(t,t^\prime)$.
\smallskip  

In the case of the three aligned vacua, if their common phase is in the small time interval $(t,t^\prime)$, the rules
of eqns.\eqref{chain3}\eqref{cenphasecond} give
\begin{equation}
 (S_{t^\prime, t})_{12}=(S_{t^\prime, t})_{23}=(S_{t^\prime, t})_{13}=1,
\end{equation}   
where the first two entries correspond to single BPS soliton chains and the last one to a BPS chain of length $2$. Now
\begin{equation}
 \exp\!\begin{scriptsize}\begin{pmatrix} 0 & 1 & \tfrac{1}{2}\\ 0 & 0 & 1\\ 0 & 0 &0 \end{pmatrix}     \end{scriptsize}=
\begin{scriptsize}\begin{pmatrix} 1 & 1 & 1\\ 0 & 1 & 1\\ 0 & 0 &1 \end{pmatrix}     \end{scriptsize}\equiv S_{t^\prime,t},
\end{equation} 
and $(L_{t^\prime,t})_{13}$  is actually \textit{required} to be $\tfrac{1}{2}$ in order to reproduce the \textit{integral} amplitude we computed in section 4. \smallskip

This analysis is easily estended to an arbitrary number of aligned vacua. Assume we have $n+1$ adiacent BPS solitons $(a_k,a_{k+1})$, all having multiplicity $1$, which are perfectly aligned and sorted in ascending order along the BPS ray,
\begin{equation}\label{situation}
	|Z_{a_0a_n}|=|Z_{a_0a_1}|+|Z_{a_1a_2}|+\cdots +|Z_{a_{n-1}a_n}|,
\end{equation}
while there is no physical BPS soliton connecting two vacua which are not next neighborhood in the $Z$--plane. 
Then the leading IR asymptotics of $(Q)_{a_0a_n}$ is as there was $\pm\tfrac{1}{n}$ BPS `solitons' of the total mass
$|Z_{a_0a_n}|$, that is $(L_{t^\prime,t})_{a_0a_n}=\pm\tfrac{1}{n}$. 

To see this, we write $T$ for the upper--triangular $(n+1)\times (n+1)$ matrix
\begin{equation}
 T_{ab}= \delta_{b,a+1},\qquad\quad T^{n+1}=0.
\end{equation} 
By assumption, the non--zero soliton multiplicities are $N_{a,a+1}=1$.
However, the arguments of section 4 give
\begin{multline}
 \big(S_{t^\prime, t}\big)_{ab}=\begin{cases}
                                 1 & \text{if } a\geq b\\
0 & \text{otherwise}.
                                \end{cases} \ \Leftrightarrow\ S_{t^\prime,t}=(\boldsymbol{1}-T)^{-1}= \boldsymbol{1}+T+T^2+\cdots + T^{n}\\
\Leftrightarrow\ L_{t^\prime,t}= \log(\boldsymbol{1}-T)^{-1}= \boldsymbol{1}+ T+ \frac{1}{2}T^2+\frac{1}{3}T^3+\cdots +\frac{1}{n}T^n.
\end{multline} 
which gives
\begin{equation}
 (L_{t^\prime,t})_{a_i,a_j}=\frac{1}{j-i}\qquad j>i,
\end{equation} 
which implies the claim. Notice that $(S^{-1}_{t^\prime,t})^T+S^{-1}_{t^\prime,t}$ is the Cartan matrix for the $A_n$ Lie algebra, in correspondence with the fact that this is the BPS spectrum of a model which flows in the UV to the $A_n$ minimal model.  The extension of this analysis to the case of inifnitely many collinear vacua and its application to $4d$,
$\cn =2$ supersymmetric theories will be discussed in \cite{cnv}.

\section{Evalutation in the Schroedinger picture:  Links to tt* Geometry}\label{secschoeredinger}

As a further check of the analysis in this paper (including the signs assignements) and to clarify its relation with the tt* geometry, we perform the same computation in the Schroedinger picture. 

The time--evolution kernel $S_{t,t^\prime}$, satisfies the (Euclidean) time-dependent Schroedinger equation\footnote{ One may prefer to call it the Fokker--Planck equation associated to the (time--dependent) stochastic equation \eqref{stoc}.}
\begin{equation}\label{schroedinger}
	\left(\frac{d}{dt}+H(t)\right)S_{t,t^\prime}=0,
\end{equation}
As $\lambda\rightarrow \infty$, the system remains most of the time near the ground states so, in the adiabatic approximation, one would write the evolution kernel as
\begin{equation}
	S_{t,t^\prime}= \Psi_a(t)\, A(t,t^\prime)^{ab}\: \Psi_b(t^\prime)^*,
\end{equation}
where $\Psi_a(t)$ is the $a$--th vacuum wave--function for the model with a superpotential with a (frozen) phase
$\exp(2i\pi k t/\beta)$ and the coefficients $A(t,t^\prime)^{ab}$ are suitable functions of $t$ and $t^\prime$. However, we must recall that, even adiabatically freezing the phase of $W$, we do not get the usual Schroedinger equation since, in that limit, our action differs from the standard one by a surface term
\begin{equation}
	S=S_\mathrm{standard}+2\,\lambda\,\mathrm{Re}\big(e^{2i\pi k t/\beta}\,W\big)_f-2\,\lambda\,\mathrm{Re}\big(e^{2i\pi k t/\beta}\,W\big)_i.
\end{equation}
Performing the compensating transformation on the states\footnote{This is the usual transformation mapping the Euclidean--time Schroerdinger equation into the forward Fokker--Planck equation.}, we obtain
\begin{equation}
	S_{t,t^\prime}= \Bigg(e^{-2\lambda\mathrm{Re}\big(e^{2i\pi k t/\beta}\,W_a\big)}\,\Psi_a(t)\Bigg) A(t,t^\prime)^{ab} \Bigg(e^{2\lambda\mathrm{Re}\big(e^{2i\pi k t^\prime/\beta}\,W_b\big)}\,\Psi_b(t^\prime)^*\Bigg).
\end{equation}

In the adiabatic approximation, which is valid as $\lambda\rightarrow \infty$, the Schroedinger equation \eqref{schroedinger} reduces to a linear differential equation for the coefficient matrix $A(t,t^\prime)^{ab}$. To get the equation satisfied by this matrix, we observe that the tt* equations imply
\begin{equation}
	\frac{d}{dt}\Psi_a = \frac{2\pi i k}{\beta}Q_{ab}\Psi_b+\cdots
\end{equation}
 where $Q_{ab}$ is the $Q$--index \cite{fendleyetal} and $+\cdots$ stands for a state orthogonal to all vacua. The Schroedinger equation then reduces to
\begin{equation}\label{finschoerdinger}
\frac{\beta}{2\pi i k}\,\frac{d}{dt}A^{ab}=Q_{ac}\,A^{cb}+2\,\lambda\,\mathrm{Im}\Big(e^{2\pi i k\tau/\lambda\beta}\, W_a\Big) A^{ab}	
	\end{equation}
which is eqn.(4.11) of ref.\cite{classification} with\footnote{As written, eqn.\eqref{finschoerdinger} differs from eqn.(4.11) of ref.\cite{classification} by exponentially small terms in the limit $\lambda\rightarrow\infty$. However, a more precise asymptotic analysis will match the subleading corrections too. To avoid misunderstanding, we stress that the $\beta$ of the present paper and the $\beta$ of ref.\cite{classification} should not be identified.}
\begin{align}
	x&= \exp\big(2\pi i k t/\beta\big)\\
\beta&=\lambda.
\end{align}
Then the IR monodromy computed with our present SQM system is the same as the monodromy computed by the tt* connection. In particular, in the extreme IR limit, the reduced time evolution kernel $A(t,t^\prime)^{ab}$  is an integral matrix taking values in the Lie groups discussed in section 6 of \cite{ttstar}. This implies that the sign rule is as in eqn.\eqref{signsrule2}. Indeed, $Q_{ab}$ is an element of the Lie algebra $\mathfrak{so}(n)$, and hence the elementary BPS multiplicity matrix $N_{ab}$, which is proportional to the leading IR contribution to $Q_{ab}$, must be real antisymmetric. In the language of Kontsevich--Soibelman \cite{KS} this property corresponds to the symmetry under the Cartan involution of $SL(n,\mathbb{Z})$,
\begin{displaymath}
 g\mapsto (g^{-1})^T.
\end{displaymath}


\end{document}